\documentclass[12pt,a4paper,final]{iopart}

\usepackage{ulem} 
\usepackage{feynmp} 
\usepackage{color}
\usepackage{multicol}
\usepackage{bm}       
\usepackage{iopams}  
\usepackage{graphicx}
\usepackage[breaklinks=true,colorlinks=true,linkcolor=blue,urlcolor=blue,citecolor=blue]{hyperref}

\begin{document}

\title{Towards a self-consistent dynamical nuclear model}

\author{X. Roca-Maza$^{1,2}$, Y.F. Niu$^{2,3}$, G. Col\`o$^{1,2}$, P. F. Bortignon$^{1,2}$}

\address{$^{1}$Dipartimento di Fisica, Universit\`a degli Studi di Milano, Via Celoria 16, I-20133 Milano, Italy\\ 
$^2$INFN, Sezione di Milano, Via Celoria 16, I-20133 Milano, Italy\\
$^3$ELI-NP, Horia Hulubei National Institute for Physics and Nuclear Engineering,
30 Reactorului Street, RO-077125, Bucharest-Magurele, Romania}

\eads{\mailto{xavier.roca.maza@mi.infn.it}}

\begin{abstract}
Density Functional Theory (DFT) is a powerful and accurate tool exploited in Nuclear Physics to investigate the ground-state and some collective properties of nuclei along the whole nuclear chart. Models based on DFT are, however, not suitable for the description of single-particle dynamics in nuclei. Following the field theoretical approach by A. Bohr and B. R. Mottelson to describe nuclear interactions between single-particle and vibrational degrees of freedom, we have undertaken important steps to build a microscopic dynamic nuclear model. In connection to that, one important issue that needs to be better understood is the renormalization of the effective interaction in the particle-vibration approach. One possible way to renormalize the interaction is the so called {\it subtraction method}. In this contribution we will implement the {\it subtraction method} for the first time in our model and study its consequences. 
\end{abstract}

\pacs{21.60.Jz,24.30Cz}

\maketitle
\section{Introduction}
\label{intro}

In our research group, we are interested in developing a microscopic nuclear structure model that can be eventually applied also to nuclear reactions. Such an achievement, would solve a long standing problem in theoretical nuclear physics \cite{nazarewicz2016}. 

We have recently developed two different energy density functionals (EDFs) \cite{ddmed,sami} based on the Density Functional Theory \cite{HK,KS2,engel2011}. In nuclear physics EDFs are commonly derived in an approximate way from a phenomenological interaction solved via the Hartree (H) or Hartree-Fock (HF) approximations \cite{bender2003,vretenar2005}. Such type of models are very helpful in understanding different nuclear properties such as nuclear masses and sizes as well as the excitation energy of nuclear collective states such as giant resonances. In the latter case a minimal extension of the theory is needed. That is, one needs to perturb the ground state density according to the oscillation mode under study. For that purpose, one may apply the small amplitude limit of the Time Dependent H or HF approximations (TDHF) which coincides with the Linear Response Theory (LRT) or the Random Phase Approximation (RPA) \cite{ringschuck, nakatsukasa2007, colo13}. We have devoted our efforts to the detailed study of such observables providing, in some cases, a useful theoretical guidance \cite{bortignonbook,centelles09,roca-maza2011,roca-maza2012,liang2012,roca-maza2013,roca-maza2013a,roca-maza2015,cao2015}.

Approaches to describe the nucleus based on the nucleon-nucleon interaction in the vacuum are successful in their predictions of some properties in light and medium mass nuclei but face computational limitations in the description of heavy systems and high-lying excited states \cite{dickhoff2004, lee2009, bogner2010, barrett2013, hagen2014, carlson2015, shen2016}. Instead, EDFs approaches do not suffer from such a limitation. Nevertheless, nuclear EDFs based on static effective potentials are not suitable for the description of single-particle dynamics in nuclei. For example the fragmentation of single-particle states \cite{bernard1980, colo2010} and their finite half-life are unequivocal finger prints. Following the field theoretical approach by A. Bohr and B. R. Mottelson \cite{BM,BMbook,BMvol} to describe nuclear interactions between single-particle and vibrational degrees of freedom, we have undertaken important steps to build a microscopic dynamic nuclear model. The Milano group has been traditionally working on such an idea via the implementation of the so called Particle Vibration Coupling (PVC) model \cite{bortignon1977, bertsch1983, colo1994}, yet different physical and technical difficulties need to be faced \cite{moghrabi2010, moghrabi2012, brenna2014, yang2016}. One of the most important drawbacks is the correct treatment of the renormalization of the interaction. By renormalization, we mean to cure the divergences whenever they appear due to the nature of the effective interaction employed and/or the determination of the renormalized parameters consistently with the adopted many-body scheme. 

In more detail, one can solve the nuclear effective Hamiltonian using perturbation theory up to first order (Hartee-Fock) and find a static solution where the nuclear field is just an average static field. The consequence is that nucleons are predicted to be frozen in their quantum states and single-particle dynamics are, thus, not realistic. To solve this problem one needs, for example, to go beyond the Hartree-Fock approximation, that is, up to higher orders in perturbation theory \cite{baldo2015}. If summing up some specific type of diagrams --those supposed to provide the largest contributions-- that correspond to collective low-energy vibrations, one can recover, ultimately, a PVC model.
Our recent efforts to understand the renormalization problem lay on the bases of a simplified model that corresponds to the lowest-order approximation on a perturbation series expansion of a microscopic PVC approach \cite{moghrabi2010, moghrabi2012, brenna2014, yang2016}. In other words, we aim at building a PVC model in a more systematic way than those adopted so far and that possibly allow us to formally and reliably treat the renormalization of the interaction vertices at all levels. 

Existing implementations of the PVC approach, based on relativistic and non-relativistic frameworks are available in the literature \cite{colo2010,bertsch1983,baldo2015,niu2012, mizuyama2012a,litvinova2013,niu2014,niu2016,lyutorovich2015}. The PVC approach has been shown to describe to a good extent the width of giant resonances, not satisfactorily explained within the TDHF, LRT or RPA. This feature is crucial to reliably estimate the beta-decay half-life of a nucleus \cite{yifei2015} or the branching ratios for $\gamma$ \cite{brenna2012} or neutron decays \cite{colo1994}. It also allows for an estimation of the dependence of the effective mass with energy and momentum \cite{mahaux1985}, or in other words, for a more realistic optical potential \cite{mizuyama2012,blanchon2015}, the essential ingredient in any nuclear reaction calculation. Based on such a successful experience, it is timely to continue the efforts of our group in which both single-particle and vibrational (phonon) degrees of freedom are taken into account and consistently calculated within the same microscopic interaction, by overcoming the difficulties related with the renormalization. 

Our strategy is to work on different fronts in order to tackle the different problems in both the nuclear effective interactions used and the many-body techniques employed. As mentioned, one important issue that needs to be understood and solved is the renormalization of the effective interaction in approximations beyond EDFs (BEDF). In this contribution we will implement the so called {\it subtraction method}\footnote{The subtraction method has been devised as a way to extend the stability condition of the RPA equations to theories beyond such an approach, e.g., second RPA (SRPA) \cite{rowe}.} \cite{tselyaev2007, tselyaev2013} to our PVC model for the first time. Such a method have been previously introduced by other groups \cite{litvinova2007, gambacurta2015, lyutorovich2015,lyutorovich2016} in order to avoid double counting when going BEDF. It is, however, not yet demonstrated that such a procedure properly renormalizes the theory.

Finally, it is important to mention that all these studies are the perfect complement to the experimental activities in our group \cite{carbone10,bocchi2016} and world wide. With the advent of new Rare Ion Beam Facilities \cite{blumenfeld2013}, the experimental investigation of proton- and neutron-rich unstable nuclei has become possible. Nuclear theory should cover and provide reliable predictions for the properties of this unexplored area of the nuclear chart, not to mention the relevance of a deep understanding of the structure of new super-heavy elements \cite{litvinova2011, oganessian2015}. 

\section{Formalism}
\label{method}

In this section, we will briefly describe the bases of our formalism paying attention to the underlying physical assumptions and refer the reader to the references herein for technical details.

\begin{figure}[h!]
\hspace{3cm}
\begin{fmffile}{hf}
\begin{fmfgraph}(150,75)
\fmfleft{i1,i2}
\fmfright{o}
\fmf{fermion}{i1,v1,i2}
\fmf{phantom}{o,o1}
\fmf{fermion,right,tension=0.4}{o1,v2,o1}
\fmf{dashes}{v1,v2}
\end{fmfgraph}
\begin{fmfgraph}(150,75)
\fmfleft{i1,i2}
\fmfright{o1,o2}
\fmf{fermion}{i1,v1}
\fmf{phantom}{v1,i2}
\fmf{fermion}{v2,o2}
\fmf{phantom}{o1,v2}
\fmf{dashes}{v1,v2}
\fmf{fermion,right=0.5,tension=0.2}{v1,v2}
\end{fmfgraph}
\end{fmffile}
\caption{Diagrammatic representation for the HF self-energy. On the left it is shown the direct term $\langle 1 2\vert V\vert 1 2\rangle$ and on the right the exchange term $\langle 1 2\vert V\vert 2 1\rangle$.}\label{fig1}
\end{figure}
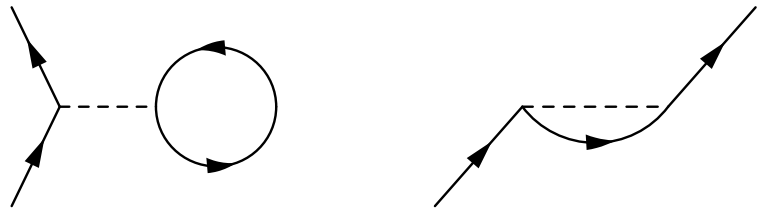

In general, the nuclear Hamiltonian can be written as $\mathcal{H} = T + V$ where $T$ represents the kinetic energy and $V$ the two-body and three-body --or density dependent two-body-- effective interaction between nucleons. Adding and subtracting an auxiliary one-body potential $U$ allows us to formulate the problem in terms of a non-interacting part $\mathcal{H}_0\equiv T + U$ that corresponds to the so called HF Hamiltonian if the auxiliary potential is defined as the ground state expectation value of $V$ on a Slater determinant, plus $V-U$ that vanishes by construction within the HF approximation. That is, the solution of $\mathcal{H}_0$ coincides with that of $\mathcal{H}$ in first order perturbation theory. The diagrammatic representation for the HF self-energy can be seen in Fig.\ref{fig1}. In what follows, we will denote as $\vert i\rangle$ the set of occupied and unoccupied HF states with energy $\varepsilon_i$. 

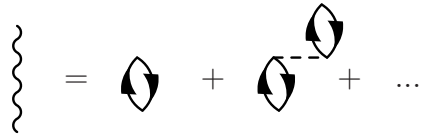
\begin{figure}[h!]
\hspace{6cm}
\begin{fmffile}{rpa}
\parbox{5mm}{
\begin{fmfgraph}(5,40)
\fmfleft{i1,i2}
\fmf{photon,tension=0}{i1,i2}
\end{fmfgraph}
}
=\hspace{0.5cm}
\parbox{7mm}{
\begin{fmfgraph}(20,40)
\fmfleft{i1,i2,i3}
\fmf{fermion,left=0.5,tension=0.2}{i2,i1}
\fmf{fermion,left=0.5,tension=0.2}{i1,i2}
\end{fmfgraph}
}
+\hspace{0.5cm}
\parbox{7mm}{
\begin{fmfgraph}(20,40)
\fmfleft{i1,i2,i3}
\fmfright{o1,o2,o3}
\fmf{fermion,left=0.5,tension=0.2}{i2,i1}
\fmf{fermion,left=0.5,tension=0.2}{i1,i2}
\fmf{dashes}{i2,o2}
\fmf{fermion,left=0.5,tension=0.2}{o3,o2}
\fmf{fermion,left=0.5,tension=0.2}{o2,o3}
\fmf{phantom,left=0.5,tension=0.2}{o2,o1}
\fmf{phantom,left=0.5,tension=0.2}{o1,o2}
\end{fmfgraph}
}
+\hspace{0.3cm} ...
\end{fmffile}
\caption{Diagrammatic representation of the infinite series of the RPA approximation. The infinite sum of bubbles represent a vibration or phonon state.}\label{fig2}
\end{figure}

Considering $V-U$ in higher order perturbation theory, that is, adopting as the unperturbed Hamiltonian $\mathcal{H}_0$, the contribution of $U$ will be exactly zero in all cases\footnote{This is because in perturbation theory, matrix elements of $V-U$ appearing in higher order contributions to the energy or wave function can be always written in terms of the unperturbed bases.}. Hence, standard perturbation theory techniques can be applied directly to the potential $V$ for higher order calculations. However, to apply such a strategy is very complicated from the technical point of view and the implementation of higher order contributions may not entail, in general, a clear physical interpretation. Alternatively, one may select the most relevant diagrams that should play a clear physical role and sum them up to infinite order --if possible. Hence, it is crucial to understand the relevant degrees of freedom of the problem under consideration and to connect them with given terms in the many-body expansion. This is the case of the vibrational degrees of freedom or phonons in nuclei\footnote{Also rotational degrees of freedom play an important role but we will not discuss this here.}. Collective low-energy nuclear vibrations constitute one of the major actors in generating, when coupled to single-particle degrees of freedom, the fragmentation of the latters that give rise to the so called spreading width observed in nuclei. At larger energies, collective motion in nuclei give rise to a prominent dynamic effect: giant resonances \cite{BMbook,bortignonbook,harakeh2001}. They are super-positions of particle-hole excitations. In particular, phonons are built from a very specific type of diagram summed up to infinity --this corresponds to solve the RPA equations. Such a series of diagrams that include a particle (unoccupied state) hole (occupied state) excitation (1p-1h), named bubble diagrams, are represented in Fig.\ref{fig2}.

Therefore, it is customary to work within a subspace $\mathcal{Q}_1$ that contains all nuclear 1p-1h configurations built with the single-particle states $\vert i\rangle$, eigenfunctions of the HF Hamiltonian, where the RPA --in a discrete space in our case-- will correspond to solve the initial Hamiltonian projected in the $\mathcal{Q}_1$ subspace $\mathcal{Q}_1\mathcal{H}\mathcal{Q}_1$. The RPA states built in this way are linear combinations of 1p-1h (HF) states that represent, as mentioned, a nuclear vibration or phonon. A discretized RPA calculation may display a spread of different excitation peaks producing a broadening of a Giant Resonance at contrast with the ideal situation in which there is a clear single and collective peak. Such an effect depends on the intensity of the residual interaction that particle-hole configurations feel as well as the density of the unperturbed 1p-1h states around a given excitation energy. This effect is well known in the literature and it is named Landau damping. In addition, it is worth mentioning that if continuum is discretized, the RPA states are stationary states with no spreading width. The reason for this is that the RPA do not introduce an energy dependence in the self-energy.

\begin{figure}[h!]
\hspace{3cm}
\begin{fmffile}{spread}
\parbox{20mm}{
\begin{fmfgraph}(30,75)
\fmfleft{i1,i2,i3,i4}
\fmfright{o1,o2,o3,o4}
\fmf{photon,right=1,tension=0.4}{i2,i3}
\fmf{fermion}{i1,i4}
\fmf{fermion}{o4,o1}
\end{fmfgraph}
}
\hspace{1cm}
\parbox{20mm}{
\begin{fmfgraph}(30,75)
\fmfleft{i1,i2,i3,i4}
\fmfright{o1,o2,o3,o4}
\fmf{photon,left=1,tension=0.4}{o2,o3}
\fmf{fermion}{i1,i4}
\fmf{fermion}{o4,o1}
\end{fmfgraph}
}
\hspace{1cm}
\parbox{20mm}{
\begin{fmfgraph}(30,75)
\fmfleft{i1,i2,i3,i4}
\fmfright{o1,o2,o3,o4}
\fmf{photon,tension=0.4}{i2,o3}
\fmf{fermion}{i1,i4}
\fmf{fermion}{o4,o1}
\end{fmfgraph}
}
\hspace{1cm}
\parbox{20mm}{
\begin{fmfgraph}(30,75)
\fmfleft{i1,i2,i3,i4}
\fmfright{o1,o2,o3,o4}
\fmf{photon,tension=0.4}{i3,o2}
\fmf{fermion}{i1,i4}
\fmf{fermion}{o4,o1}
\end{fmfgraph}
}
\end{fmffile}
\caption{Diagrammatic representation of the terms contributing to the spreading width in the PVC approach.}\label{fig3}
\end{figure}
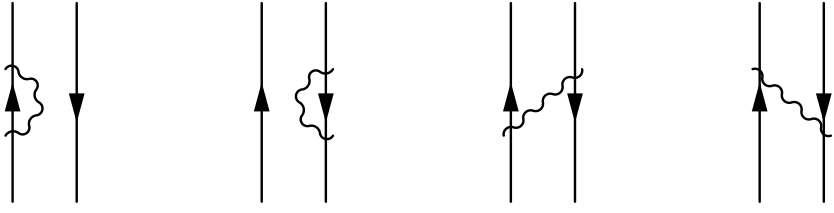

In general, the coupling of nuclear vibrational states with single-particle states\footnote{For consistency within the adopted many-body scheme, it is important to note that when coupling a phonon state with a particle state the one bubble diagram have to be subtracted from the self-energy\cite{baldo2015} (see Fig.2 of the same reference). The reason is that the RPA phonon contains the contribution of the one bubble diagram. Such diagram has two equivalent fermionic lines and, thus, for symmetry reasons a factor half has to be taken into account. The way of implementing this symmetry is just by subtracting the one bubble diagram. The latter considerations do not apply for higher order bubble diagramms.} consents the transfer of energy from one degree of freedom to another. This allows for the rearrangement of the internal degrees of freedom giving rise to a damping via the so called spreading width ($\Gamma^{\downarrow}$). A PVC approach provides also a more realistic description --when compared to the HF or RPA results-- of the emission of a $\gamma$ ray and the escape of nucleons which contribtues to the so called escape width ($\Gamma^\uparrow$). In Fig.\ref{fig3} we show the diagrammatic representation of the terms contributing to the spreading width in the PVC approach.

As in Ref.\cite{colo1994}, in order to model the escape and spreading widths, we define two additional sub-spaces. The first one $\mathcal{P}$ is made of holes plus unbound HF states which have positive energy and that we construct to be orthogonal to the bound occupied and unoccupied HF states $\vert i\rangle$ by solving $\mathcal{H}_0$ for those additional scattering states labeled as $\vert s\rangle$. The second subspace $\mathcal{Q}_2$ is made of 1p-1h excitations $\vert ph\rangle$ coupled to a phonon state labeled as $\vert N\rangle$. Note that by construction $\mathcal{Q}_1\mathcal{Q}_2=0$ since $\langle ph\vert N\otimes ph\rangle$ will be always zero. Finally, we define the subspace $\mathcal{Q}=\mathcal{Q}_1+\mathcal{Q}_2$. The projectors $\mathcal{Q}$ and $\mathcal{P}$ cover the full model space: $\mathcal{Q+P}=1$ and $\mathcal{Q}\mathcal{P}=0$; and all the projectors fulfill the usual conditions: $\mathcal{P}^2=\mathcal{P}$, $\mathcal{Q}^2=\mathcal{Q}$ and $\mathcal{Q}_i^2=\mathcal{Q}_i$. 

The Green's function formalism gives a natural representation of the response of a quantum mechanical system to a given perturbation. The dynamical many-body Green's function $\mathcal{G}$ of an interacting system described by a Hamiltonian $\mathcal{H}$ evaluated at a given energy $\omega$, is a solution of the operator equation 
\begin{equation}
(\omega-\mathcal{H}-i\epsilon)\mathcal{G}(\omega) = 1 \ .
\label{g}
\end{equation}
Within our model, the Green's function $\mathcal{G}(\omega)$ can be split in different terms using the defined sub-spaces as 
\begin{equation}
\mathcal{G}= \mathcal{Q}\mathcal{G}\mathcal{Q} + \mathcal{Q}\mathcal{G}\mathcal{P} + \mathcal{P}\mathcal{G}\mathcal{Q} + \mathcal{P}\mathcal{G}\mathcal{P} \ .
\end{equation}

Our interest is to provide a microscopic and realistic description of the dynamics in nuclei. We therefore choose to project the Hamiltonian in the $\mathcal{Q}_1$ subspace taking into account in an approximate way the effects of the continuum $\mathcal{P}$ and of more complex configurations $\mathcal{Q}_2$. 

Using the properties of the sub-spaces defined here and following a method very similar to that of Ref.\cite{yoshida1983}, we manipulate Eq.(\ref{g}) by first sandwiching it with $\mathcal{Q}_1$, 
\begin{equation}
\mathcal{Q}_1 (\omega-\mathcal{H}-i\epsilon)\mathcal{Q}_1\cdot\mathcal{Q}_1\mathcal{G}\mathcal{Q}_1 + \mathcal{Q}_1\mathcal{H}\mathcal{Q}_2\cdot\mathcal{Q}_2\mathcal{G}\mathcal{Q}_1+\mathcal{Q}_1\mathcal{H}\mathcal{P}\cdot\mathcal{P}\mathcal{G}\mathcal{Q}_1 = \mathcal{Q}_1 ;
\label{q1}
\end{equation}
then, similarly, we sandwich it with $\mathcal{P}$ from the left and $\mathcal{Q}_1$ from the right and, finally, with $\mathcal{Q}_2$ from the left and $\mathcal{Q}_1$ from the right. From the last two equations, analogous to Eq.(\ref{q1}), we find an expression for  $\mathcal{Q}_2\mathcal{G}\mathcal{Q}_1$ and $\mathcal{P}\mathcal{G}\mathcal{Q}_1$ in terms of $\mathcal{Q}_1\mathcal{G}\mathcal{Q}_1$ that we insert into Eq.(\ref{q1}). Putting all that together one arrives at the expression, 
\begin{equation}
(\omega-\mathcal{H}_{\mathcal{Q}_1}-i\epsilon)\mathcal{Q}_1\mathcal{G}(\omega)\mathcal{Q}_1 = \mathcal{Q}_1 \ ,
\end{equation}
where $\mathcal{H}$ projected in the $\mathcal{Q}_1$ subspace is (cf. \ref{appendix})
\begin{eqnarray}
\mathcal{H}_{\mathcal{Q}_1} &=& \mathcal{Q}_1\mathcal{H}\mathcal{Q}_1 \nonumber\\ 
&+&\mathcal{Q}_1\mathcal{H}\mathcal{P}\frac{1}{\mathcal{P}(\omega-\mathcal{H}-i\epsilon)\mathcal{P}}\mathcal{P}\mathcal{H}\mathcal{Q}_1 \nonumber\\
&+& \mathcal{Q}_1\mathcal{H}\mathcal{Q}_2\frac{1}{\mathcal{Q}_2(\omega-\mathcal{H}-i\epsilon)\mathcal{Q}_2}\mathcal{Q}_2\mathcal{H}\mathcal{Q}_1 
\nonumber\\  
&+&\mathcal{Q}_1\mathcal{H}\mathcal{P}\frac{1}{\mathcal{P}(\omega-\mathcal{H}-i\epsilon)\mathcal{P}}\mathcal{P}\mathcal{H}\mathcal{Q}_2\frac{1}{\mathcal{Q}_2(\omega-\mathcal{H}-i\epsilon)\mathcal{Q}_2}\mathcal{Q}_2\mathcal{H}\mathcal{Q}_1\nonumber\\
&+&\mathcal{Q}_1\mathcal{H}\mathcal{Q}_2\frac{1}{\mathcal{Q}_2(\omega-\mathcal{H}-i\epsilon)\mathcal{Q}_2}\mathcal{Q}_2\mathcal{H}\mathcal{P}\frac{1}{\mathcal{P}(\omega-\mathcal{H}-i\epsilon)\mathcal{P}}\mathcal{P}\mathcal{H}\mathcal{Q}_1 \nonumber\\
&+& {\rm higher~order~terms} \ .
\label{hq1.0}
\end{eqnarray}
Assuming that continuum states are unaffected by collective vibrations $\mathcal{P}\mathcal{H}\mathcal{Q}_2=\mathcal{Q}_2\mathcal{H}\mathcal{P}=0$ (see also \ref{appendix}), one will recover the expression in Eq.(24) of Ref.\cite{yoshida1983}. Such an approximation is justified as follows. Matrix elements of $\mathcal{P}V\mathcal{Q}$ are expected to be very small due to the short range of the interaction: the overlapping between continuum states and bound states or phonon states will occur in space regions where the former wave functions are small. Therefore, $\mathcal{P}\mathcal{H}\mathcal{Q}\approx \mathcal{P}\mathcal{H}_0\mathcal{Q}$ where $\mathcal{P}\mathcal{H}_0\mathcal{Q}_2=0$ by construction. In Eq.(\ref{hq1.0}), the first term in the right hand side of the equation corresponds to the RPA solution, $\mathcal{Q}_1\mathcal{H}\mathcal{Q}_1$; the second term corresponds to excite a bound particle or hole state to the continuum, its propagation on top of the non-interacting (HF) potential for states with positive energy and its deexcitation to a bound state again. This term, commonly labeled as $\mathcal{W}^\uparrow$, will produce the escape width previously discussed. The fourth term corresponds to the coupling of a particle or hole state with more complex configurations represented by $\mathcal{Q}_2$, its propagation on top of the potential, and the reabsortion of this complex state into a particle or a hole. This term, commonly labeled as $\mathcal{W}^\downarrow$, will produce the spreading width previously discussed. The last two terms are higher order contributions to $\mathcal{W}^\uparrow$ and $\mathcal{W}^\downarrow$ that account for the effects of nuclear vibrations into continuum states. Those terms are neglected in our calculations due to the reasons discussed above. 

Hence, we define within our approximations,      
\begin{eqnarray}
\mathcal{W}^\uparrow(\omega) &\equiv& \mathcal{Q}_1\mathcal{H}\mathcal{P}\frac{1}{\mathcal{P}(\omega-\mathcal{H}-i\epsilon)\mathcal{P}}\mathcal{P}\mathcal{H}\mathcal{Q}_1 
\label{dw} 
\\
\mathcal{W}^\downarrow(\omega) &\equiv& \mathcal{Q}_1\mathcal{H}\mathcal{Q}_2\frac{1}{\mathcal{Q}_2(\omega-\mathcal{H}-i\epsilon)\mathcal{Q}_2}\mathcal{Q}_2\mathcal{H}\mathcal{Q}_1 \ .  
\label{sw}
\end{eqnarray}
For the escape term, within our previous approximations, one can take advantage of $\mathcal{Q}\mathcal{H}\mathcal{P}\approx\mathcal{Q}\mathcal{H}_0\mathcal{P}$ and consistently employ the Green's function solution of $(\omega-\mathcal{H}_0-i\epsilon)\mathcal{G}_0 = 1$. Therefore, similar to Eq.(29) of Ref.\cite{yoshida1983}, one can write it as
\begin{eqnarray}
\mathcal{W}^\uparrow(\omega) &=& \mathcal{Q}_1(\omega-\mathcal{H}_0-i\epsilon)\mathcal{Q}_1 -\frac{1}{\mathcal{Q}_1\mathcal{G}_0\mathcal{Q}_1} \nonumber\\
&=& \omega - \mathcal{Q}_1\mathcal{H}_0\mathcal{Q}_1 -\frac{1}{\mathcal{Q}_1\mathcal{G}_0\mathcal{Q}_1} \ .
\end{eqnarray}
Provided a sufficiently large basis $\vert i\rangle$ is employed, the accuracy of this procedure is comparable with exact continuum RPA calculations \cite{giai1987}. 

In what follows, we define some useful and related quantities. The observed spectrum of a nucleus excited by an external field $\mathcal{F}$ is described by the nuclear polarization propagator or dynamic polarizability $\Pi(\omega)$. It corresponds to the double convolution with $\mathcal{F}$ of the propagator or response function, i.e. in our specific case $\mathcal{G}_{\mathcal{Q}_1}(\omega)\equiv\mathcal{Q}_1\mathcal{G}(\omega)\mathcal{Q}_1=(\omega-\mathcal{H}_{\mathcal{Q}_1}(\omega)+i\epsilon)^{-1}$, that is
\begin{equation}
\Pi(\omega) = \langle 0\vert \mathcal{F}^\dag\frac{1}{\omega-\mathcal{H}_{\mathcal{Q}_1}(\omega)+i\epsilon}\mathcal{F}\vert 0\rangle \ .
\label{resp}
\end{equation}
The strength function is defined as 
\begin{equation}
\mathcal{S}(\omega) = -\frac{1}{\pi}\Im\left[\Pi(\omega)\right] \ . 
\end{equation}
Finally, of special interest are some of the moments of the strength function since they are subject to fulfill some existing sum rules. The $k-$moment of the strength function is defined as
\begin{equation}
m_k=\int_0^\infty \omega^{k} \mathcal{S}(\omega) d\omega \ ,
\end{equation}
where $m_0$ corresponds to the so called Non Energy Weighted Sum Rule (NEWSR), $m_1$ corresponds to the so called Energy Weighted Sum Rule (EWSR) and $m_{-1}$ to the Inverse Energy Weighted Sum Rule (IEWSR). The latter is proportional to the static limit of the dynamic polarizability $\Pi(\omega=0)=-2m_{-1}$.  

\subsection{The subtraction method}
\label{subtraction}

Commonly in nuclear physics, the determination of EDF parameters is done at the H or HF levels as previously explained. Obviously, effective theories that go beyond the H or HF approaches should be refitted to experimental data in order to avoid double-counting. That is, a renormalization of the model parameters is compulsory with respect to those determined within the H or HF approaches. The parameters will change their value since physical many-body terms beyond the H or HF approximations are now explicitly included. The purpose of the {\it subtraction method} \cite{tselyaev2007, tselyaev2013} is to provide a recipe for the renormalization of the effective interaction within the adopted model scheme that avoids a refitting of the parameters. We will see that such a method is suitable to renormalize the expectation value of one body operators only. Therefore, if this is equal to --or one proper way to-- a refit of the interaction, is a question that would depend on the nature of the studied (fitted) observables. Another related but different issue is the renormalization of the divergences that appear in BEDF theories when zero-range effective interactions are adopted (see for example \cite{moghrabi2010,brenna2014} and Sec.\ref{reno} below). 

Apart from the renormalization of the interaction, the {\it subtraction method} proposed in Ref.\cite{tselyaev2013} has been devised to keep the stability condition of an RPA-like matrix when going beyond the RPA. This should be one of the most important points to justify such a procedure. The stability condition guarantees real eigenvalues and implies that the Slater determinant on which the RPA-like matrices are based must be a minimum of the energy. On this regard, we note that our theoretical method allows us to write the equations to be solved as an energy dependent RPA-like matrix (cf. Eq.(2.5) in Ref.\cite{colo1994}).  


In what follows, we review the underlying idea of the {\it subtraction method}. For technical details we refer the reader to the original reference \cite{tselyaev2013}. In general, the dynamic polarizability $\Pi(\omega)$ follows the equation 
\begin{eqnarray}
\Pi(\omega) = \Pi_0(\omega) + \Pi_0(\omega)W(\omega)\Pi(\omega)
\label{prop}
\end{eqnarray}
where $W(\omega)$ represents in general an induced effective interaction and $\Pi_0(\omega)$ the dynamic polarizability at the level of approximation that one wants to improve. That is, if the RPA approach is adopted (see Fig.\ref{fig2}), $\Pi_0=\Pi_{\rm HF}$ and $W$ is the so called particle-hole (or residual) interaction that is defined as 
\begin{eqnarray}
W_{\rm RPA}(1,2) \equiv \frac{\delta^2E[\rho]}{\delta \rho_1 \delta \rho_2} 
\end{eqnarray}
where $E[\rho]$ represents the EDF of choice.

Now, assuming that we are dealing with the exact nuclear EDF $E[\rho]$ and that it can be derived from a Hamiltonian, DFT ensures that $E[\rho; \lambda] = E[\rho] + \lambda\mathcal{O}$ will also describe exactly the ground state of the perturbed system if $\mathcal{O}$ is a one-body operator. Such an observation implies that the expectation value of any one-body operator calculated using the ground-state wave function solution of the constrained calculation, i.e. solution of $\tilde{\mathcal{H}}=\mathcal{H}+\lambda\mathcal{O}$, is also exact. Via the dielectric theorem \cite{bohigas1979}, which establishes that $m_{-1}=\frac{1}{2} \partial^2_\lambda \langle\lambda\vert\mathcal{H}\vert\lambda\rangle\vert_{\lambda=0}$, one should conclude that $m_{-1}$ should be conserved in BEDF calculations as compared to its value calculated within the exact EDF. In this case, it is shown that $m_{-1}$ should be conserved but it might be that other features related to the response function require the same treatment.   

One of the possible realizations that conserve the value of $m_{-1}$ in BEDF approaches is as suggested in \cite{tselyaev2013}. In this reference it is shown that recovering the static limit of the dynamic polarizability is equivalent to recover the RPA response function $\mathcal{G}_{\rm RPA}$ in BEDF approaches. This would imply in practice to modify Eq.(\ref{prop}) as follows:
\begin{eqnarray}
\Pi_{\rm BEDF}(\omega) = \Pi_{\rm RPA} + \Pi_{\rm RPA}\left[W_{\rm BEDF}(\omega)-W_{\rm BEDF}(\omega=0)\right]\Pi(\omega)_{\rm BEDF}  
\end{eqnarray}
since now it is ensured that $\Pi_{\rm BEDF}(\omega=0) = \Pi_{\rm RPA}$ and, therefore, $\mathcal{G}_{\rm BEDF}(\omega=0)=\mathcal{G}_{\rm RPA}$ (cf. Eq.\ref{resp}). In our specific case, the induced effective interaction will be $W_{\rm PVC}(\omega)\equiv \mathcal{W}^\uparrow(\omega)+\mathcal{W}^\downarrow(\omega)-\mathcal{W}^\downarrow(\omega=0)$.

Hence, the {\it subtraction method} directly impacts on many properties of excited modes in nuclei althought it should restore the EDF values for the expectation value of one-body operators due to the redefinition (renormalization) of $W(\omega)$. Therefore, to study how such a method performs in practice for those cases in which we know that the result should be conserved in BEDF calculations with respect EDF calculations is of prominent importance.

\subsection{Sum rules}

The Thouless theorem \cite{thouless1961} states that the EWSR calculated within the RPA approach is equal to the HF expectation value of the double commutator $\frac{1}{2}[\mathcal{F},[\mathcal{H},\mathcal{F}^\dag]]$, where $\mathcal{F}$ represents the external field that perturbs the nuclear ground state. In Ref.\cite{yannouleas1987} it was proven that the EWSR in second RPA (SRPA) \cite{rowe,wambach1990} is also equal to the double commutator calculated within the HF ground state and, therefore, also to the RPA value. Such a result was derived without implementing the {\it subtraction method}. Actually, this proof is valid for both the full SRPA and SRPA calculations where the 2$p$-2$h$ subspace has been eliminated by introducing an energy dependent effective interaction in the 1$p$-1$h$ subspace \cite{wambach1990} (see below). 
As a matter of fact, RPA as well as the latter proof are based on the quasiboson approximation, where the expectation value of the operators are calculated within the uncorrelated HF ground state instead of the more consistent treatment that would consider the correlated RPA (or SRPA) ground state \cite{catara1996}. On this regard, the {\it subtraction method} should provide a correlated wave function for the ground state that should give the same results, when applied to the calculation of the expectation value of any one body operator, as the HF expectation value of the same operator. This is because, the redefinition of the induced effective interaction in the static limit ($\omega=0$). Due to these considerations, when the result of the double commutator $[\mathcal{F},[\mathcal{H},\mathcal{F}^\dag]]$ is a one-body operator, the EWSR calculated within the SRPA --and possibly also the PVC-- method with and without subtraction should not differ much in practice with the RPA result.

Regarding the IEWSR or static polarizability $\Pi(\omega=0)=-2m_{-1}$: from Ref.\cite{wambach1990} it is easy to understand the amount by which $W$ should be corrected in SRPA to recover the same value for the IEWSR found in RPA [cf. Eq.(2.87)]. It actually coincides with the {\it subtraction method} of Ref.\cite{tselyaev2013}. An easy way to see that is as follows. Starting from the (full) SRPA matrix [see for example Eq.(2.48) in Ref.\cite{wambach1990}] where 1$p$-1$h$ and 2$p$-2$h$ terms are assumed to interact, one can easily rearrange rows and columns such that the resulting matrix separates within different sub-spaces that enclose the 1$p$-1$h$ ($\mathcal{Q}_1$) and 2$p$-2$h$ ($\mathcal{Q}_2^\prime$) sub-spaces separated in blocks along the diagonal while outside the diagonal there appear the interaction terms between both sub-spaces. By using the technique described above \cite{yoshida1983}, one may project the original Hamiltonian $\mathcal{H}$ into the $\mathcal{Q}_1$ subspace taking into account the effects of the $\mathcal{Q}_2^\prime$ subspace perturbatively. For that the only assumption is to neglect the residual interaction in the $\mathcal{Q}_2^\prime$ subspace. Having done that, the induced effective interaction can be modified by an energy dependent term
\begin{equation}
W_{\rm SRPA}(\omega) = W_{\rm RPA} + \mathcal{Q}_1\mathcal{H}\mathcal{Q}_2^\prime\frac{1}{\mathcal{Q}_2^\prime(\omega-\mathcal{H}-i\epsilon)\mathcal{Q}_2^\prime}\mathcal{Q}_2^\prime\mathcal{H}\mathcal{Q}_1 \ .
\label{srpa}
\end{equation}   
The latter correction written within the standard SRPA matrix formulation can be seen in Eq.(2.69) of Ref.\cite{wambach1990}. By inspecting now the expression for the $m_{-1}$ in SRPA (see also Eq.(2.87) of Ref.\cite{wambach1990}) and imposing that $m_{-1}$(SRPA)=$m_{-1}$(RPA) or equivalently $\Pi_{\rm SRPA}(\omega=0)$=$\Pi_{\rm RPA}$, one easily realizes from Eq.(\ref{resp}) that there is an extra term that one will need to subtract in order to fulfill the latter equality. Such term actually coincides with the one proposed by the {\it subtraction method}. Therefore, for sufficiently weak perturbations, the energy dependent SRPA reduction from the full SRPA should be accurate and the subtraction method proposed by Tselyaev \cite{tselyaev2013} should properly work and conserve the value of the static polarizability in such an extension BEDF. 

We will now analyze $m_1$ and $m_{-1}$ from a different perspective. Consider that the ground state density is perturbed by an external (one-body) field $\lambda \mathcal{F}$. Changes in the expectation value of the Hamiltonian $\mathcal{H}$ can be written as,
\begin{equation}  
\delta\langle\mathcal{H}\rangle_{\mathcal{F}} = \lambda^2\sum_{n\neq 0}\frac{\vert\langle n\vert\mathcal{F}\vert 0\rangle\vert^2}{E_n} + \mathcal{O}(\lambda^3) = \lambda^2 m_{-1} + \mathcal{O}(\lambda^3)
\end{equation}
where standard perturbation theory has been applied (i.e. $\vert n\rangle$ and $E_n$ represents an excited state and corresponding energy of the system).
In other terms, 
\begin{equation}  
m_{-1} = \frac{1}{2}\frac{\partial^2\langle\mathcal{H}\rangle_{\mathcal{F}}}{\partial \lambda^2}\Big\vert_{\lambda=0}
\end{equation}
which is nothing but the dielectric theorem \cite{bohigas1979} previously introduced. If we consider the case in which $\mathcal{F}$ is an isoscalar and velocity independent operator and define the operator $\tilde{\mathcal{F}}\equiv i[\mathcal{H},\mathcal{F}] = i[T,\mathcal{F}]$ where $T$ is the kinetic energy, we can calculate the change in the expectation value of the Hamiltonian when perturbed by $\tilde{\mathcal{F}}$ and find
\begin{equation}  
\delta\langle\mathcal{H}\rangle_{\tilde{\mathcal{F}}} = \lambda^2\sum_{n\neq 0} E_n \vert\langle n\vert\mathcal{F}\vert 0\rangle\vert^2 + \mathcal{O}(\lambda^3) = \lambda^2 m_1 + \mathcal{O}(\lambda^3)
\end{equation}
This observation \cite{bohigas1979} allows one to state that for some specific operators (excitation modes) {\it the Thouless theorem for $\mathcal{F}$ is equivalent to the dielectric theorem applied to $\tilde{\mathcal{F}}$}, 
\begin{equation}  
m_1 = \frac{1}{2}\frac{\partial^2\langle\mathcal{H}\rangle_{\tilde{\mathcal{F}}}}{\partial \lambda^2}\Big\vert_{\lambda=0} = \frac{1}{2}\langle 0\vert [\mathcal{F},[\mathcal{H},\mathcal{F}]]\vert 0\rangle \ ,
\label{m1}
\end{equation}
provided that the corresponding quantities are calculated consistently within the same approximation. Hence, we see again that whenever $\langle 0\vert [\mathcal{F},[\mathcal{H},\mathcal{F}]]\vert 0\rangle_{\rm BEDF}\approx \langle 0\vert [\mathcal{F},[\mathcal{H},\mathcal{F}]]\vert 0\rangle_{\rm HF}$ the EWSR of some special excitation modes, should be conserved when going BEDF.

As a final remark, it has been also shown in Ref.\cite{wambach1990} that the NEWSR $m_0$ within the SRPA approach coincides with the same quantity calculated within the RPA approach whenever a subtraction is not implemented. To check this feature together with the results for the $m_1$ and $m_{-1}$ within the PVC approach might shed some light into the connection between the sum rules calculated at different levels of approximation in the adopted many-body scheme and on the renormalization of the particle-vibration approach.

\subsection{Ultraviolet divergences}
\label{reno}

So far, the considerations here do not take into account the renormalization of the ultraviolate divergences arising from BEDF models if based on zero-range effective interactions \cite{moghrabi2010}, such as the widely used Skyrme as well as part of the Gogny interaction \cite{decharge1980} or the so-called point-coupling relativistic models \cite{nikolaus1992}. Interestingly, ultraviolate divergences seem to be avoided by the {\it subtraction method}. Hence, in this context, the {\it subtraction method} can be regarded as a practical recipe that should be employed with caution since it is yet to be demonstrated that it propely renormalizes the theory. The reabsortion of the ultraviolate divergence by the {\it subtraction method} is as follows. In general, one can write in second order perturbation theory
\begin{equation}
W(\omega;1,2)-W(\omega=0;1,2) = \sum_{n}\frac{\langle 1\vert V\vert n\rangle\langle n\vert V\vert 2\rangle}{\omega-\omega_n} + \sum_{n}\frac{\langle 1\vert V\vert n\rangle\langle n\vert V\vert 2\rangle}{\omega_n}
\end{equation}
where 1 and 2 indicate two different nuclear states and $n$ is a complete bases of nuclear intermediate states that connect the initial and final states via the effective interaction $V$. As it is evident from the last equation, for $\omega_n \rightarrow \infty$ the divergence is canceled. 


\section{Results}
\label{results}

As discussed, one of the important test grounds for the {\it subtraction method} is the analysis of some specific sum rules. With the questions raised in the previous section in mind, we will study the $m_0$ (NEWSR), $m_1$ (EWSR), $m_{-1}$ (IEWSR) and the centroid energies $m_1/m_0$ and $\sqrt{m_1 / m_{-1}}$ for the isoscalar monopole $\mathcal{F}_{0}^{\rm IS}=\sum_{i=1}^{A}r_i^2Y_{00}(\hat{r}_i)$ and quadrupole $\mathcal{F}_{2}^{\rm IS}=\sum_M\sum_{i=1}^{A}r_i^2Y_{2M}(\hat{r}_i)$ modes of excitation in the test case of ${}^{16}$O. This will be in analogy with Ref.\cite{gambacurta2015} where the same cases have been studied within the SRPA approach. The selection of these one-body, isoscalar and velocity independent external fields is motivated by the discussions made in the previous section in connection with the calculation of $m_1$ and $m_{-1}$. In order to assess the systematics on our results, we have adopted three different non-relativistic effective interactions SLy5 \cite{sly5}, SkM* \cite{skm} and SAMi \cite{sami} of common use in nuclear physics.

\begin{figure}[t!]                    
\includegraphics[width=1\linewidth,clip=true]{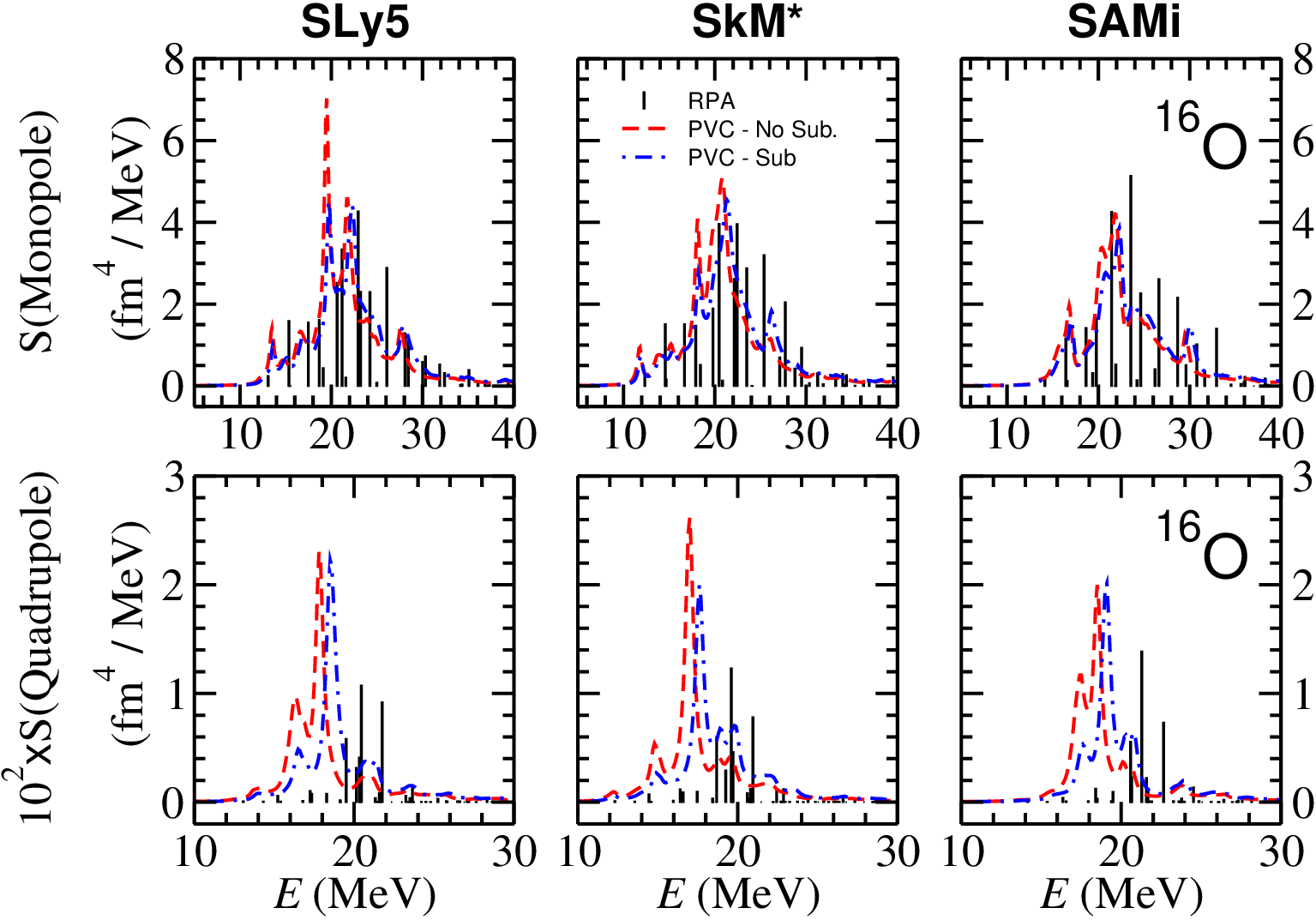}
\caption{Results for the monopole (upper panels) and quadrupole (lower panels) response in ${}^{16}$O as predicted by three Skyrme interactions: SLy5 \cite{sly5} (left panels), SkM* \cite{skm} (middle panels) and SAMi \cite{sami} (right panels). In grey bars, the RPA response is depicted. In dashed red lines PVC results without subtraction and in dash-point lines with subtraction are shown.} 
\label{fig4} 
\end{figure}

In addition, we will show our results for the isoscalar monople response in ${}^{208}$Pb due to its relevance in the determination of the nuclear matter incompressibility and of the isoscalar quadrupole response in ${}^{208}$Pb since it is connected to the value of the effective mass at the Fermi surface. We will also comment on our results for the low-lying $2_1^+$ state in ${}^{208}$Pb that has been argued in Ref.\cite{lyutorovich2016} not to be strongly affected by complex configurations such as the coupling to a phonon state.

The model space in our calculations is defined by a radial mesh of 200 points in steps of 0.1 fm and a maximum particle energy of 80 MeV. We have checked that doorway phonons of non-natural parity are negligible for the studied cases and, thus, we only include phonons of natural parity up to multipolarity equal to 5 with energy less than 30 MeV and absorbing a fraction of the NEWSR larger than 2\%. Such a choice gives converged results for the different moments of the strength function studied here. We use a smearing parameter of 250 keV. The full effective interaction is self-consistently kept in all vertices at all levels of approximation studied here. Our model has been tested also for the well known Gamow-Teller resonance \cite{niu2014,yifei2015} without accounting for the subtraction and contemporarily to our present work in Ref.\cite{niu2016} including the subtraction.

\subsection{Sum rules}

The EWSR for the isoscalar monopole and quadrupole responses in ${}^{16}$O has been seen to change in SRPA calculations with respect to the double commutator sum rule calculated at the HF level when the subtraction method is adopted (cf. Fig.13 in Ref.\cite{gambacurta2015}). In the same work the EWSR is conserved when the subtraction is not implemented. This  is in agreement with Refs.\cite{yannouleas1987,wambach1990}, where it was shown that the EWSR is conserved for SRPA calculations without subtraction. As a matter of fact, the changes produced by the subtraction method on the prediction of the EWSR in the SRPA calculations presented in Ref.\cite{gambacurta2015} are not large ($\lesssim$10\%).

\begin{figure}[t!]                    
\includegraphics[width=\linewidth,clip=true]{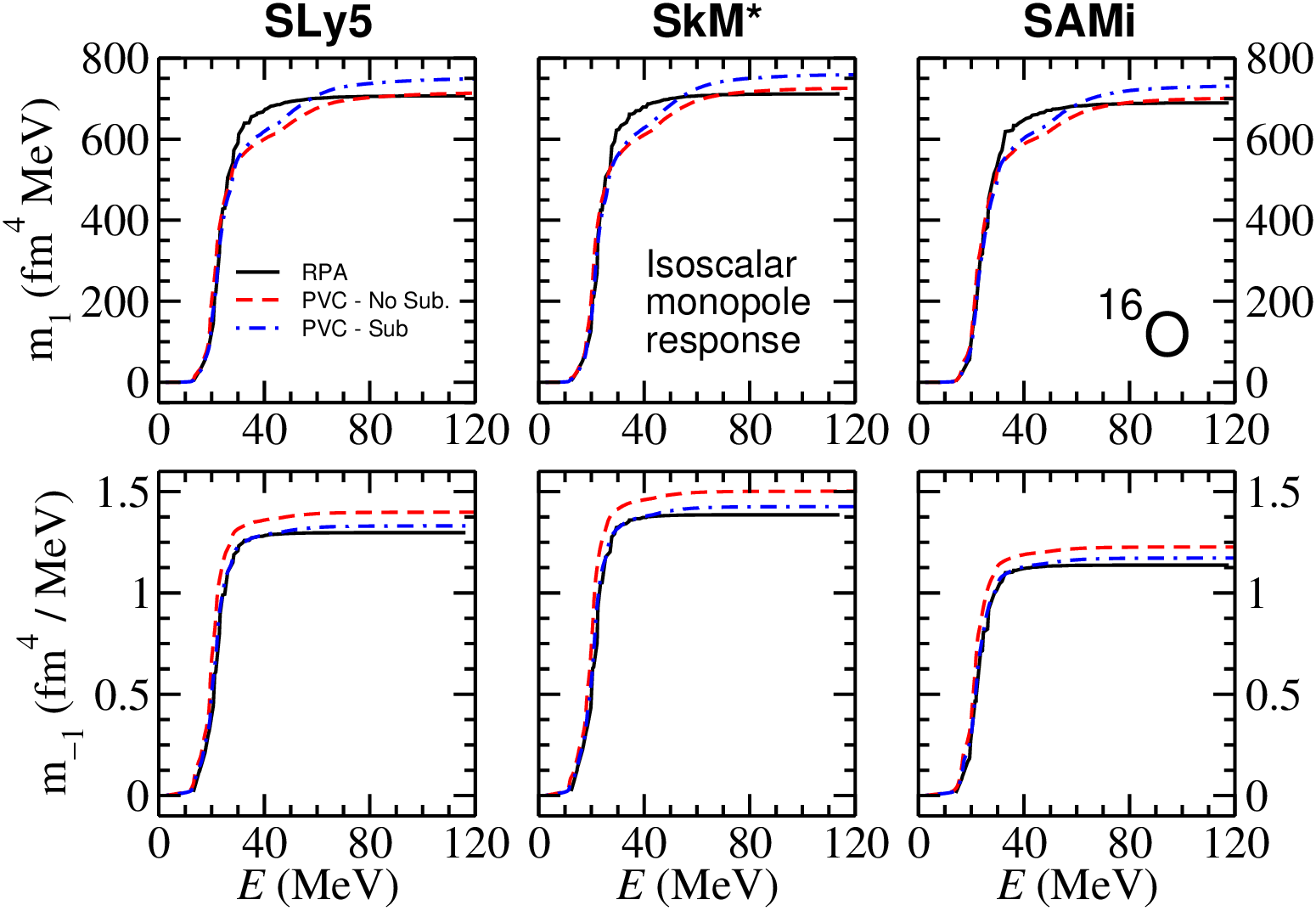}
\caption{Cumulative sums for the EWSR (upper panels) and IEWSR (lower panels) for the monopole response in ${}^{16}$O as predicted by three Skyrme interactions: SLy5 \cite{sly5} (left panels), SkM* \cite{skm} (middle panels) and SAMi \cite{sami} (right panels). In black lines the RPA results are depicted while PVC results are shown by dashed red lines without subtraction and by dash-point lines with subtraction.} 
\label{fig5} 
\end{figure}

In Fig.\ref{fig4}, we show our PVC results for the monopole (upper panels) and quadrupole (lower panels) response in ${}^{16}$O as predicted by three Skyrme interactions: SLy5 (left panels), SkM* (middle panels) and SAMi (right panels). In grey bars, the RPA response is depicted. In dashed red lines PVC results without subtraction and in dash-point lines with subtraction are shown. For the case of the monopole response, PVC correlations do not affect qualitatively the strength function in the giant resonance region and, consistently, the subtraction method has a little impact. Such a small PVC effect is well understood from theory (see Sec.IV.B in \cite{bertsch1983}). The first (second) and third (fourth) diagrams reading from the left in Fig.\ref{fig3} correspond to matrix elements of the type particle-particle (hole-hole) squared and particle-particle and hole-hole mixed that differ by a geometrical factor (6-$j$ symbol, cf. Eq.(A12) in Ref.\cite{colo1994}): this matrix elements are essentially equal and with opposite sign in the case of $L=0$ excitation modes.     

For the quadrupole response instead, we see from Fig. \ref{fig4} that the PVC results qualitatively modify the strength function with respect to the RPA results by producing a shift in energy of the giant resonance peak. This is because the real part of the self-energy or, equivalently, the effective mass has been strongly modified by the PVC \cite{bertsch1968}. Such a change is expected since we know that the effective mass ($m^*$) at the Fermi surface ($m^*/m \approx 1$ where $m$ is the bare nucleon mass) is not well described within EDF models ($m^*/m \approx 0.7$) and it is corrected in the right direction by the PVC approach \cite{mahaux1985}. We will come back to this point for the case of ${}^{208}$Pb.

\begin{figure}[t!]                    
\includegraphics[width=1\linewidth,clip=true]{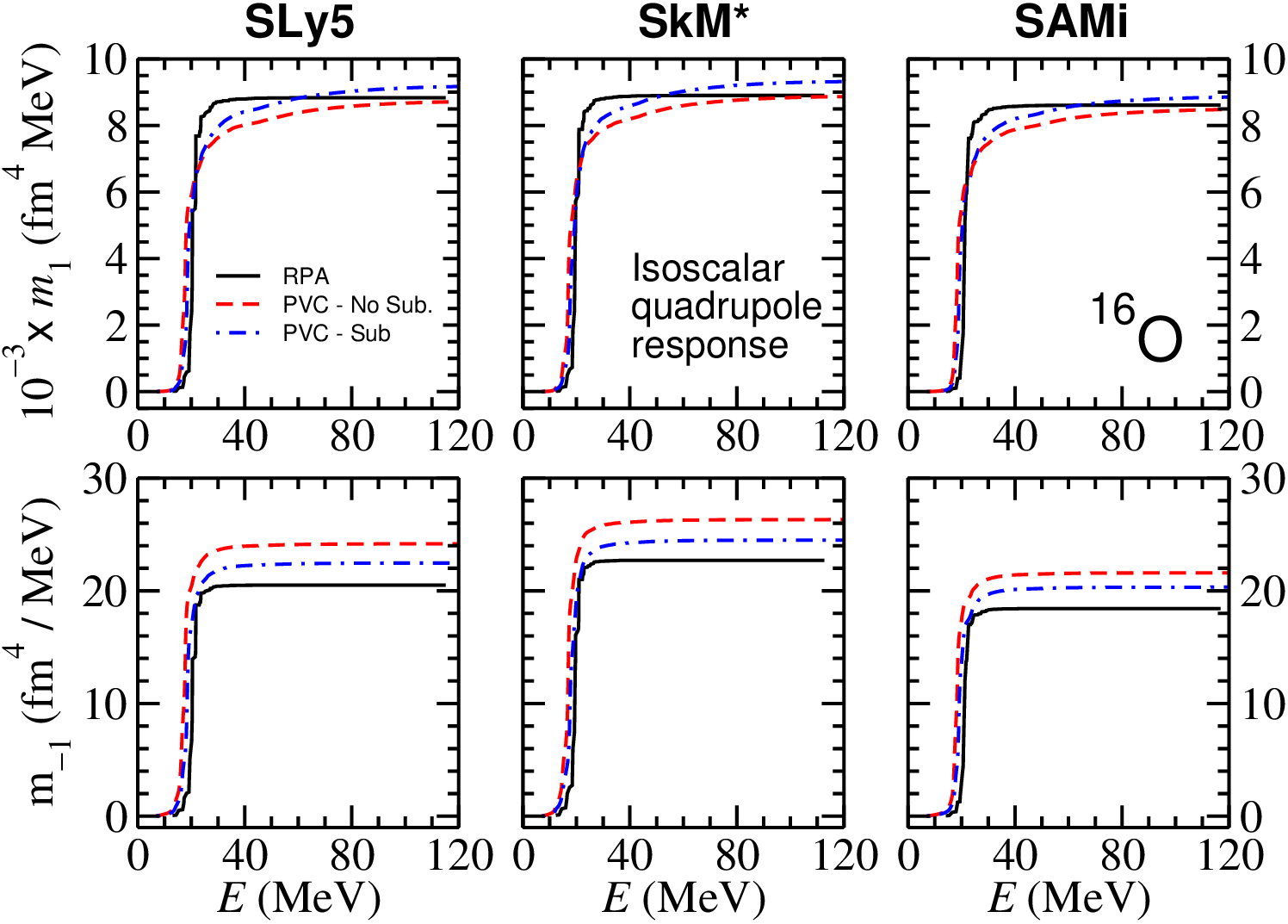}
\caption{Cumulative sums for the EWSR (upper panels) and IEWSR (lower panels) for the quadrupole response in ${}^{16}$O as predicted by three Skyrme interactions: SLy5 \cite{sly5} (left panels), SkM* \cite{skm} (middle panels) and SAMi \cite{sami} (right panels). In black lines the RPA results are depicted while PVC results are shown by dashed red lines without subtraction and by dash-point lines with subtraction.} 
\label{fig6} 
\end{figure}

The differences on the predicted sum rules for the monopole case can be clearly seen in Fig.\ref{fig5} (and Table \ref{tab1}) where cumulative sums for $m_1$ (upper panels) and $m_{-1}$ (lower panels) are displayed following the same color and line-type code as in Fig.\ref{fig4}. In this figure, it is clear that the EWSR is recovered within the PVC approach with no subtraction while PVC results with subtraction slightly overestimate $m_1$ by 6-7\% (cf. Table \ref{tab1}). This is in agreement with Ref.\cite{yannouleas1987,wambach1990}. Regarding the IEWSR, it is increased by 8\% in the PVC predictions with no subtraction while it is almost perfectly restored when the subtraction is implemented in agreement with what we expected from the previous considerations and with Ref.\cite{tselyaev2013}. In Table \ref{tab1}, the converged values for the monopole sum rules are shown. In addition to the quantitative information discussed above on these results, we also notice that our RPA results fully exhaust the double commutator sum rule; that the NEWSR in the PVC calculations with and without subtraction are (almost) equal to that of the RPA; that, consequently, the centroid energy $m_1/m_0$ is barely shifted by the PVC calculations without subtraction while it is shifted upwards by about 5\% (1 MeV) if the subtraction is implemented; and that, consequently, the PVC predictions for the centroid energy $\sqrt{m_1/m_{-1}}$ show the same trends although its value is slightly closer to the RPA when the subtraction is implemeted ($\lesssim$2\%).

\begin{table}[t!]
\caption{\label{tab1} NEWSR ($m_0$) in fm$^4$, EWSR ($m_1$) in fm$^4$ MeV, IEWSR ($m_{-1}$) in fm$^4$/MeV and the centroid energies $m_1/m_0$ and $\sqrt{m_1/m_{-1}}$ in MeV for the monopole response in ${}^{16}$O. D.C. stands for the double commutator sum rule  calculated within the HF ground state. The \% related to the PVC results are referred to RPA and those of the RPA to the D.C..} 
\begin{center}
\begin{tabular}{rcccccccc}
 \hline\hline
Force &Sum rule& D.C. & RPA & [\%] & PVC (No Sub.) & [\%]& PVC (Sub.) & [\%]\\
 \hline
SLy5   &$m_{ 0}$&    &29.5& &29.9& 101&29.8& 101\\
       &$m_{ 1}$& 706&707 & 100& 713 & 101& 749  & 106\\
       &$m_{-1}$&    &1.30&    & 1.40& 108& 1.33 & 102\\
       &$m_{ 1}/m_0$&    &24.0& &23.9&100 &25.1& 105\\
       &$\sqrt{m_1/m_{-1}}$& &23.3& &22.6&96&23.7&100\\
 \hline
SkM$^*$&$m_{ 0}$&    &30.5& &31.2& 102&31.1& 102\\
       &$m_{ 1}$& 712&712 & 100& 726 & 102& 759  & 107\\
       &$m_{-1}$&    &1.39&    & 1.50& 108& 1.43 & 103\\
       &$m_{ 1}/m_0$&    &23.3& &23.3&100 &24.4& 105\\
       &$\sqrt{m_1/m_{-1}}$& &22.7& & 22.0&97 & 23.1&102 \\
 \hline
SAMi   &$m_{ 0}$&    &27.3& &27.8& 102&27.8& 102\\
       &$m_{ 1}$& 688&689 & 100& 701 & 102& 731  & 106\\ 
       &$m_{-1}$&    &1.14&    & 1.23& 108& 1.17 & 103\\
       &$m_{ 1}/m_0$&    &25.3& &25.2&100 &26.3& 104\\
       &$\sqrt{m_1/m_{-1}}$& &24.6& & 23.9&97 & 25.0& 102\\
 \hline\hline
\end{tabular}
\end{center}
\end{table}

The differences on the predicted sum rules for the quadrupole case can be seen instead in Fig.\ref{fig6} (and Table \ref{tab2}) where cumulative sums for $m_1$ (upper panels) and $m_{-1}$ (lower panels) are displayed following the same color and line-type code as in Fig.\ref{fig4}. In this figure, it is clear that the EWSR is recovered within the PVC approach with no subtraction while PVC results with subtraction slightly overestimate $m_1$ by 3-5\% (cf. Table \ref{tab2}). This is again in agreement with Refs.\cite{yannouleas1987,wambach1990} and with our previous discussions. Regarding the IEWSR, it is increased by 16-20\% in the PVC predictions with no subtraction while it is only partially restored within 10\% when the subtraction is implemented. Therefore, the {\it subtraction method} in our model does not perfectly work in this case although the correction is sizeable and in the right direction. In Table \ref{tab2}, the converged values for the quadrupole sum rules are shown. In addition to the quantitative information discussed above on these results, we should notice that our RPA results fully exhaust the double commutator sum rule; that the NEWSR in the PVC calculations with and without subtraction are overestimated by $\lesssim 5\%$ with respect to the RPA values; and that, consequently, the centroid energies $m_1/m_0$ and $\sqrt{m_1/m_{-1}}$ are barely shifted by the PVC calculations with subtraction while they are shifted downwards by about 5-10\% ($\lesssim$ 2 MeV) if the subtraction is not implemented. Hence, our results are not conclusive yet although the {\it subtraction method} gives reasonable results --within 10\% accuracy-- for the studied sum rules. The reason for that discrepancy is actually a measure of the accuracy of the adopted approximations which are essentially two:  i) the $\mathcal{Q}_2$ subspace is assumed to be made of non-interacting states; and ii) we do not correct for the small --but present-- contributions to Eq.(\ref{sw}) that violate the Pauli exclusion principle. Work to solve these issues should be addressed in the future.

\begin{table}[t!]
\caption{\label{tab2}  NEWSR ($m_0$) in fm$^4$, EWSR ($m_1$) in fm$^4$ MeV, IEWSR ($m_{-1}$) in fm$^4$/MeV and the centroid energies $m_1/m_0$ and $\sqrt{m_1/m_{-1}}$ in MeV for the quadrupole response in ${}^{16}$O. D.C. stands for the double commutator sum rule  calculated within the HF ground state. The \% related to the PVC results are referred to RPA and those of the RPA to the D.C..} 
\begin{center}
\begin{tabular}{rcccccccc}
 \hline\hline
Force &Sum rule& D.C. & RPA & [\%] & PVC (No Sub.) & [\%]& PVC (Sub.) & [\%]\\
 \hline
SLy5   &$m_{ 0}$&    &423& &443&105 &439& 104\\
       &$m_{ 1}$& 8829 & 8836 & 100 &8712 & 99&9169 & 104\\
       &$m_{-1}$&      & 20.50&     &24.17&120&22.46& 110\\
       &$m_{ 1}/m_0$&    &20.9& &19.7&94 &20.9& 100\\
       &$\sqrt{m_1/m_{-1}}$& &20.8 & & 19.0 &91 & 20.2 &97 \\
 \hline
SkM$^*$&$m_{ 0}$&    &447& &467&104 &462& 103\\
       &$m_{ 1}$& 8902 & 8902 & 100 &8866 &100&9317 & 105\\
       &$m_{-1}$&      & 22.71&     &26.31&116&24.50&108\\
       &$m_{ 1}/m_0$&    &19.9& &19.0& 95&20.2& 102\\
       &$\sqrt{m_1/m_{-1}}$& &19.8 & &  18.4 &93 &  19.5 & 98\\
 \hline
SAMi   &$m_{ 0}$&    &397& &415&105 &412& 104\\
       &$m_{ 1}$& 8603 & 8612 & 100 &8482 & 99&8853 & 103\\ 
       &$m_{-1}$&      & 18.43&     &21.60&117&20.32& 110\\
       &$m_{ 1}/m_0$&    &21.7& &20.4& 94&21.5& 99\\
       &$\sqrt{m_1/m_{-1}}$& &21.6 & &  19.8 & 92&  20.9 & 97\\
 \hline\hline
\end{tabular}
\end{center}
\end{table} 

\subsection{Isoscalar quadrupole response in ${}^{208}$Pb}  

\begin{figure}[t!]                    
\centerline{\includegraphics[width=0.8\linewidth,clip=true]{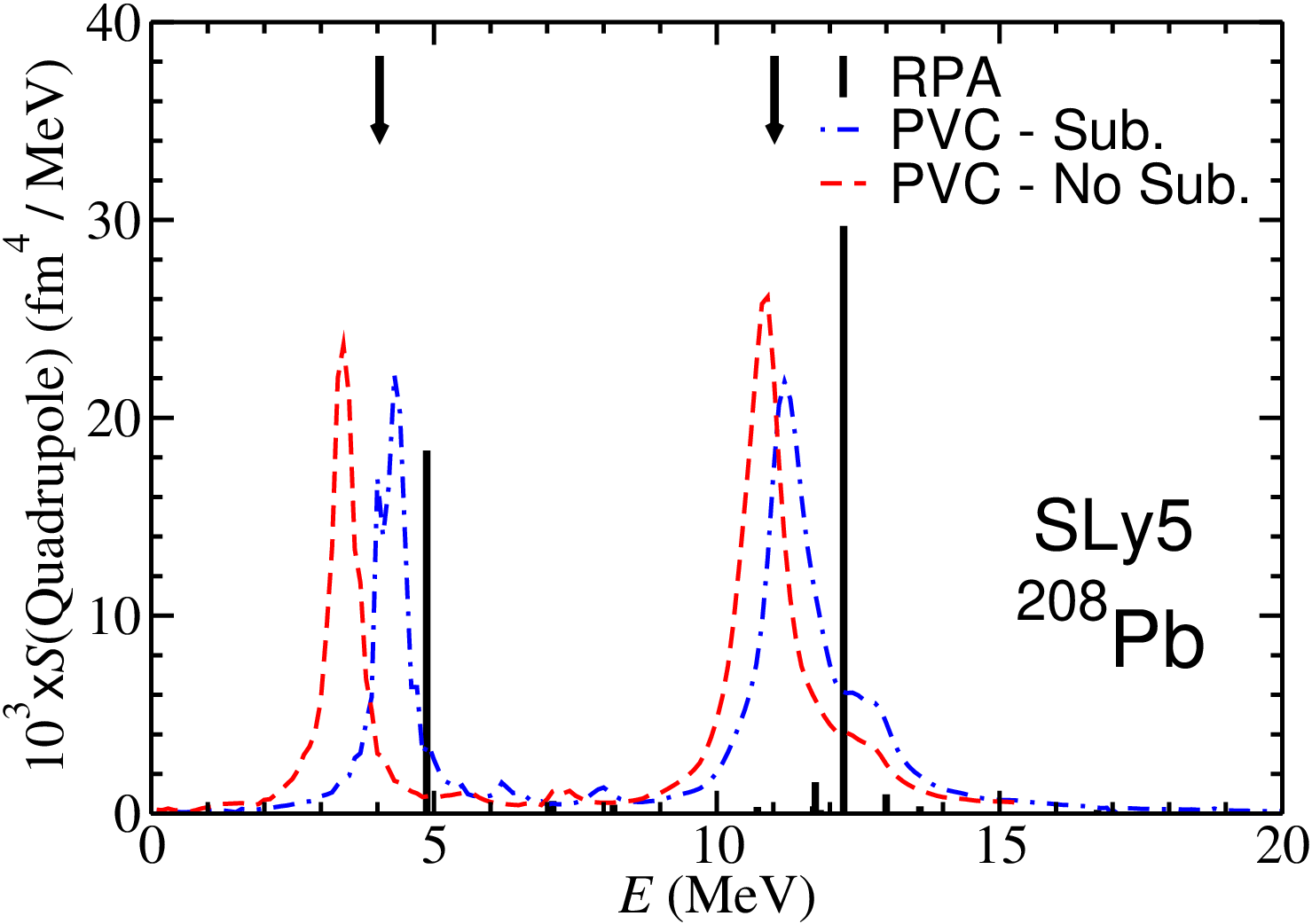}}
\caption{Isoscalar quadrupole response in ${}^{208}$Pb as predicted by SLy5. In grey bars we show the RPA response, in dashed red lines the PVC without subtraction and in dot dashed blue lines the PVC with subtraction. One black arrow indicates the position of the measured $2_1^+$ state and the other black arrow indicates the experimental centroid energy $m_1/m_0=11.0\pm 0.2$ MeV\cite{youngblood1981}.} 
\label{fig7} 
\end{figure}

Experimentally, the well known $2_1^+$ state in the low-energy isoscalar quadrupole response in ${}^{208}$Pb is at about 4 MeV exhausting a large fraction of the EWSR while the giant resonance peak is at around 11 MeV and has a width of about 3 MeV \cite{harakeh2001}. The high energy peak corresponding to the isoscalar giant quadrupole resonance is known to be related to the value of the effective mass in the vicinity of the Fermi surface ($m^*/m \sim 1$) \cite{BMbook}. Specifically, within a simple harmonic oscillator model, one may write that $E_x({\rm ISGQR}) = \sqrt{2m/m^*}\hbar\omega$ where $\hbar\omega=41 A^{-1/3}$ is the shell gap. We show in Fig.\ref{fig7} the predictions of our PVC calculations using SLy5. As grey bars we show the RPA response, in dashed red lines the PVC without subtraction and in dot dashed blue lines the PVC with subtraction. One black arrow indicates the position of the measured $2_1^+$ state and the other black arrow indicates the experimental centroid energy $m_1/m_0=11.0\pm 0.2$ MeV in the giant resonance region\cite{youngblood1981}. PVC calculations show that the energy of the $2_1^+$ state is affected by more complex configurations that produce a downshift in energy of about 2 MeV with respect to the RPA result when the subtraction is not implemented and it is only slightly shifted when the subtraction is applied giving a much realistic estimate for such state. In this regard, our results without subtraction are not satisfactory. Regarding the giant resonance peak, as it is well known, RPA calculations based on EDF with effective masses lower than the bare nucleon mass overestimates the excitation energy. As it is also well known, the PVC approach introduces an energy dependence also in the real part of the self-energy that corrects the value of the effective mass \cite{bertsch1968} such that it is more realistic and compares better with the empirical value. On this regard, it seems that the results employing the {\it subtraction method} better reproduces the excitation energy of the giant resonance. 

\subsection{Isoscalar monopole response in ${}^{208}$Pb} 

\begin{figure}[t!]                    
\centerline{\includegraphics[width=0.8\linewidth,clip=true]{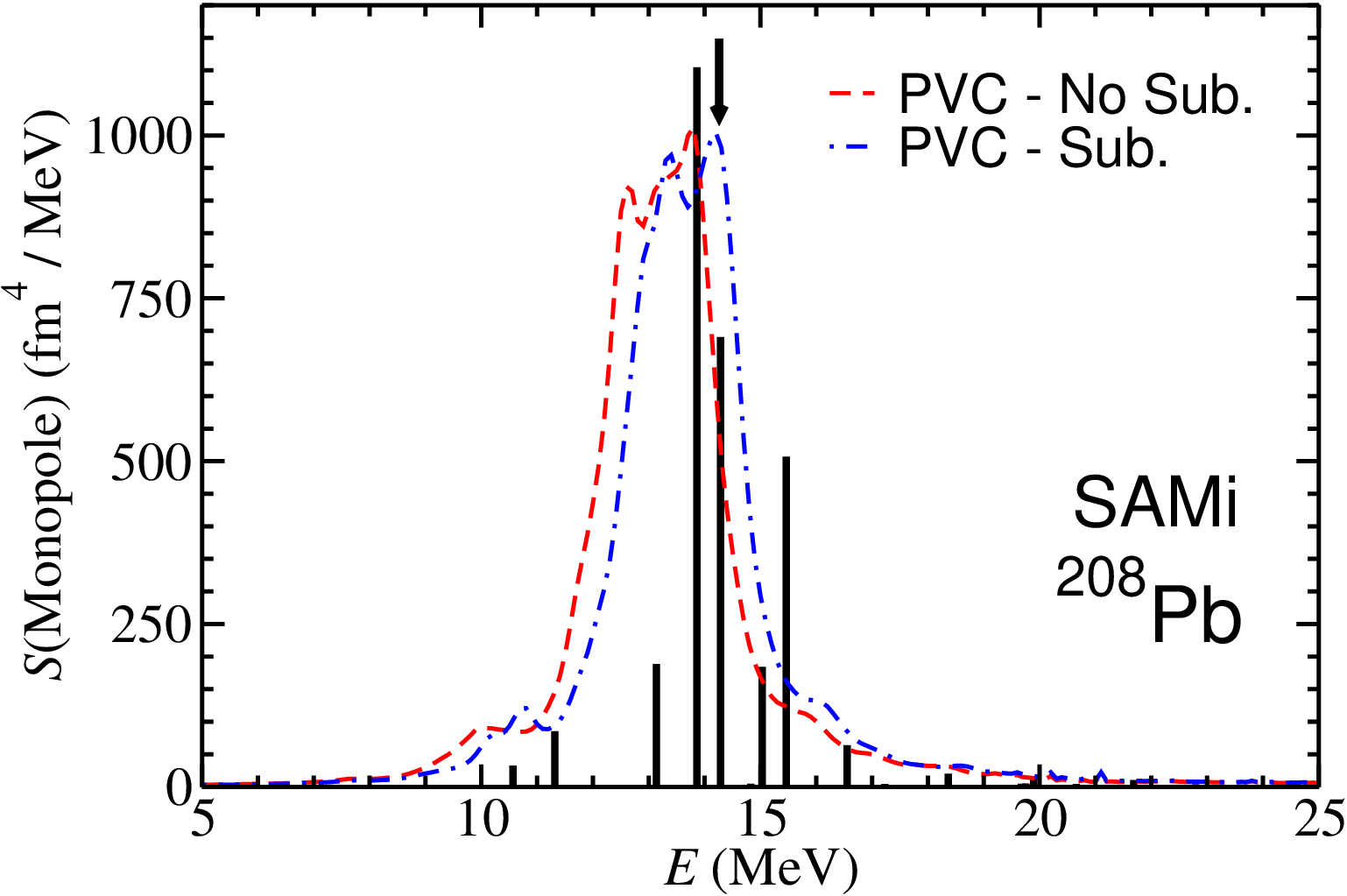}}
\caption{Isoscalar monopole response in ${}^{208}$Pb as predicted by SAMi. In grey bars we show the RPA response, in dashed red lines the PVC without subtraction and in dot dashed blue lines the PVC with subtraction. A black arrow indicates the centroid energy $m_1/m_0=14.24\pm 0.11$ MeV measured within 8 and 22 MeV \cite{youngblood1999}.} 
\label{fig8} 
\end{figure}

In Fig.\ref{fig8}, the isoscalar monopole response in ${}^{208}$Pb as predicted by the SAMi interaction is shown. In grey bars we show the RPA response while in dashed red lines the PVC without subtraction and in dot dashed blue lines the PVC with subtraction are displayed. A black arrow indicates the centroid energy $m_1/m_0=14.24\pm 0.11$ MeV measured within 8 and 22 MeV \cite{youngblood1999}. The centroid energy defined as the square root of the ratio between the EWSR and IEWSR is $\sqrt{m_1/m_{-1}}=14.18\pm 0.11$ MeV\cite{youngblood1999}. The width of the resonance has been measured to be between 2 and 3 MeV approximately (cf. Table 4.1 of Ref.\cite{harakeh2001}). In our calculations using SAMi, we find within 8 and 22 MeV that for the RPA the EWSR is 97\%, $m_1/m_0 = 13.7$ MeV and $\sqrt{m_1/m_{-1}} = 13.5$ MeV; for the PVC without subtraction  the EWSR is 91\%, $m_1/m_0 = 13.4$ MeV and $\sqrt{m_1/m_{-1}} = 13.2$ MeV; and for the PVC with subtraction the EWSR is 91\%, $m_1/m_0 = 13.7$ MeV and $\sqrt{m_1/m_{-1}} = 13.6$ MeV. The width predicted by our PVC calculations is of 2 MeV. Thus, SAMi predicts reasonable values for the excitation energy and width of this resonance both, with and without subtraction, since PVC effects are small as expected. It is well known within the RPA approach that the excitation energy of the isoscalar giant monopole resonance and the finite nucleus incompressibility $K_A$ can be related as follows \cite{khan2013}: $E_x({\rm ISGMR}) \equiv \sqrt{m_1/m_{-1}} = \sqrt{\hbar^2 K_A/(m\langle r^2\rangle)}$ where $\langle r^2\rangle$ is the mean squared radius of the nucleus. Hence, our results suggest that $K_A$ will be barely affected by PVC effects and, then, it should be reliably derived from EDF models. In the limit of $A\rightarrow\infty$ one would recover the value for the nuclear matter incompressibility, a crucial ingredient of the nuclear Equation of State that governs physical systems from the very small: the interior of a nucleus; to the very big: the interior of a neutron star.  

\section{Conclusions}
\label{conclusions}

The PVC approach presents some important points that need to be solved. In connection to that, one issue that needs to be better understood is the renormalization of the effective interaction in the particle-vibration approach. One proposed way to do so is the so called {\it subtraction method} \cite{tselyaev2013}. This method ensures that the static limit of the dynamic polarizability is conserved in BEDF approaches with respect to the EDF value. 
In addition, the {\it subtraction method} guaranties the stability condition in RPA-like theories. Both features are very important. As an example, the static dipole polarizability is being nowadays extensively studied theoretically on an EDF bases (see \cite{roca-maza2015} and references therein) and in laboratories such as the RCNP in Japan or GSI in Germany \cite{tamii2011,rossi2013,hashimoto2015}. 

In this contribution we have implemented the {\it subtraction method} for the first time in our PVC model and studied its suitability on the bases of existing sum rules applied as an example to the case of ${}^{16}$O and ${}^{208}$Pb. We have found that the {\it subtraction method} allows one to (mostly) recover in the PVC approach the value of $m_{-1}$ predicted at the RPA level but with some caveats: while this is almost exactly fulfilled in our calculations of the isoscalar monopole resonance in ${}^{16}$O, it is just approximately fulfilled for the case of the isoscalar quadrupole response in ${}^{16}$O. These results should be further and systematically investigated in the future. As discussed, also the $m_0$ and $m_1$ moments of the strength function have been subject of our studies. This is because these two quantities have been shown to be conserved by SRPA calculations with respect to their RPA counterparts when the subtraction is not implemented \cite{yannouleas1987, wambach1990}. In addition, the $m_1$ moment has been intensively studied along the years, both experimentally and theoretically \cite{bortignonbook, harakeh2001}. Within our calculations, the $m_1$ value is exactly conserved at the PVC level with respect the corresponding RPA results only if the subtraction is not implemented and slightly overestimated otherwise. 

The quantities $m_1/m_0$ and $\sqrt{m_1/m_{-1}}$ have been studied since they constitute one of the possible ways to extract the excitation energy of a given resonance. Both have been object of many studies in the past and their study should be revitalized nowadays with the advent of new Rare Ion Beam facilities that aim at measuring the excitation properties in exotic nuclei. Regarding the excitation energy $\sqrt{m_1/m_{-1}}$, we have seen that our PVC calculations almost recover the RPA value when the {\it subtraction method} is applied while it is slightly underestimated otherwise. For the centroid energy $m_1/m_0$, as a consequence of our previous results, we find that it is conserved for the monopole if the subtraction is not implemented and for the quadrupole if the subtraction is implemented. 

We have also presented our results on the isoscalar monopole and quadrupole responses in ${}^{208}$Pb and learn that the finite nucleus incompressibility is barely affected by PVC effects while the effective mass --as it was well known-- is properly corrected in our PVC calculations. In addition to that we have paid special attention to the $2_1^+$ state since it has been argued \cite{lyutorovich2016} that it should not be strongly modified by complex configurations such as the coupling to a phonon state. We find that this is not the case when the subtraction is not implemented since the energy of this peak is downshifted by about 2 MeV going far from the experimental value and that it is just slightly shifted with respect to the RPA result when the subtraction is applied, the latter giving a much realistic estimate for such an state. 

As a summary, in all cases the moments and excitation energies agree within $\lesssim$10\% with the RPA results except for the calculation of $m_{-1}$ within our PVC model without implementing the subtraction. The studied giant resonances in ${}^{208}$Pb as well as the $2_1^+$ state are in reasonable agreement with experiment by our PVC calculations if the subtraction is implemented. Our results indicate that the {\it subtraction method} renormalizes the particle-vibration approach in the good direction although a better understanding or strategy is indeed needed on this regard. Further investigations in connection with the renormalization of the particle-vibration approach are envisaged.

\appendix
\section{Effective Hamiltonian in the $\mathcal{Q}_1$ subspace}
\label{appendix}
Using the properties of the sub-spaces defined here and following a method very similar to that of Ref.\cite{yoshida1983}, we manipulate Eq.(\ref{g}) by first sandwiching it with $\mathcal{Q}_1$, 
\begin{equation}
\hspace*{-2cm}
\mathcal{Q}_1 (\omega-\mathcal{H}-i\epsilon)\mathcal{Q}_1\cdot\mathcal{Q}_1\mathcal{G}\mathcal{Q}_1 + \mathcal{Q}_1\mathcal{H}\mathcal{Q}_2\cdot\mathcal{Q}_2\mathcal{G}\mathcal{Q}_1+\mathcal{Q}_1\mathcal{H}\mathcal{P}\cdot\mathcal{P}\mathcal{G}\mathcal{Q}_1 = \mathcal{Q}_1 
\label{a1}
\end{equation}
then, similarly, we sandwich it with $\mathcal{P}$ from the left and $\mathcal{Q}_1$ from the right and, finally, by doing it with $\mathcal{Q}_2$ from the left and $\mathcal{Q}_1$ from the right. This allows us to find the system of operator equations: 
\[
\hspace*{-2cm}
   \left( {\begin{array}{cc}
            $$\mathcal{P}(\omega-\mathcal{H}-i\epsilon)\mathcal{P}$$ & $$-\mathcal{P}\mathcal{H}\mathcal{Q}_2$$ \\       
            $$-\mathcal{Q}_2\mathcal{H}\mathcal{P}$$ & $$\mathcal{Q}_2(\omega-\mathcal{H}-i\epsilon)\mathcal{Q}_2$$ \\      
           \end{array} } \right)
   \left( {\begin{array}{c}
            $$\mathcal{P}\mathcal{G}\mathcal{Q}_1$$ \\       
            $$\mathcal{Q}_2\mathcal{G}\mathcal{Q}_1$$ \\      
           \end{array} } \right) 
=
   \left( {\begin{array}{c}
            $$\mathcal{P}\mathcal{H}\mathcal{Q}_1\cdot\mathcal{Q}_1\mathcal{G}\mathcal{Q}_1$$ \\       
            $$\mathcal{Q}_2\mathcal{H}\mathcal{Q}_1\cdot\mathcal{Q}_1\mathcal{G}\mathcal{Q}_1$$ \\      
           \end{array} } \right)
\]
From these system of equations, one can find an expression for  $\mathcal{Q}_2\mathcal{G}\mathcal{Q}_1$ and $\mathcal{P}\mathcal{G}\mathcal{Q}_1$ in terms of $\mathcal{Q}_1\mathcal{G}\mathcal{Q}_1$ by inverting the matrix at the left hand side. Inserting such a solution in Eq.(\ref{a1}) one can find the final expression for $\mathcal{H}_{\mathcal{Q}_1}$ that fulfills
\begin{equation}
(\omega-\mathcal{H}_{\mathcal{Q}_1}-i\epsilon)\mathcal{Q}_1\mathcal{G}(\omega)\mathcal{Q}_1 = \mathcal{Q}_1 \ .
\end{equation}
If we assume $\mathcal{P}\mathcal{H}\mathcal{Q}_2$ and $\mathcal{Q}_2\mathcal{H}\mathcal{P}$ are negligible in the matrix at the left hand side of the system, it is easy to invert it and find an expression for $\mathcal{P}\mathcal{G}\mathcal{Q}_1$ and $\mathcal{Q}_2\mathcal{G}\mathcal{Q}_1$ as a function of $\mathcal{Q}_1\mathcal{G}\mathcal{Q}_1$ 
\[
\hspace*{-2cm}
   \left( {\begin{array}{c}
            $$\mathcal{P}\mathcal{G}\mathcal{Q}_1$$ \\       
            $$\mathcal{Q}_2\mathcal{G}\mathcal{Q}_1$$ \\      
           \end{array} } \right) 
\approx
   \left( {\begin{array}{cc}
            $$\frac{1}{\mathcal{P}(\omega-\mathcal{H}-i\epsilon)\mathcal{P}}$$ & 0 \\       
            0 & $$\frac{1}{\mathcal{Q}_2(\omega-\mathcal{H}-i\epsilon)\mathcal{Q}_2}$$ \\      
           \end{array} } \right)
   \left( {\begin{array}{c}
            $$\mathcal{P}\mathcal{H}\mathcal{Q}_1\cdot\mathcal{Q}_1\mathcal{G}\mathcal{Q}_1$$ \\       
            $$\mathcal{Q}_2\mathcal{H}\mathcal{Q}_1\cdot\mathcal{Q}_1\mathcal{G}\mathcal{Q}_1$$ \\      
           \end{array} } \right)
\]
Using the latter expressions, we recover $\mathcal{H}_{\mathcal{Q}_1}$ of Ref.\cite{yoshida1983}. If we assume instead that  $\mathcal{P}\mathcal{H}\mathcal{Q}_2$ and $\mathcal{Q}_2\mathcal{H}\mathcal{P}$ are small as compared to the diagonal terms of the same matrix, we can approximately invert the matrix in the left hand side of the equation and find 
{\small
\[
\hspace*{-4cm}
   \left( {\begin{array}{c}
            $$\mathcal{P}\mathcal{G}\mathcal{Q}_1$$ \\       
            $$\mathcal{Q}_2\mathcal{G}\mathcal{Q}_1$$ \\      
           \end{array} } \right) 
\approx\\
   \left( {\begin{array}{cc}
            $$\frac{1}{\mathcal{P}(\omega-\mathcal{H}-i\epsilon)\mathcal{P}}$$ & $$\frac{1}{\mathcal{P}(\omega-\mathcal{H}-i\epsilon)\mathcal{P}}\mathcal{P}\mathcal{H}\mathcal{Q}_2\frac{1}{\mathcal{Q}_2(\omega-\mathcal{H}-i\epsilon)\mathcal{Q}_2}$$ \\       
            $$\frac{1}{\mathcal{Q}_2(\omega-\mathcal{H}-i\epsilon)\mathcal{Q}_2}\mathcal{Q}_2\mathcal{H}\mathcal{P}\frac{1}{\mathcal{P}(\omega-\mathcal{H}-i\epsilon)\mathcal{P}}$$ & $$\frac{1}{\mathcal{Q}_2(\omega-\mathcal{H}-i\epsilon)\mathcal{Q}_2}$$ \\      
           \end{array} } \right)
   \left( {\begin{array}{c}
            $$\mathcal{P}\mathcal{H}\mathcal{Q}_1\cdot\mathcal{Q}_1\mathcal{G}\mathcal{Q}_1$$ \\       
            $$\mathcal{Q}_2\mathcal{H}\mathcal{Q}_1\cdot\mathcal{Q}_1\mathcal{G}\mathcal{Q}_1$$ \\      
           \end{array} } \right)
\]
}
Using this result, we can find a more accurate expression for $\mathcal{H}_{\mathcal{Q}_1}$ that also contain the effects of collective vibrations into the continuum states. That is [cf. Eq.(\ref{hq1.0})], 
\begin{eqnarray}
\mathcal{H}_{\mathcal{Q}_1} &\approx& \mathcal{Q}_1\mathcal{H}\mathcal{Q}_1 \nonumber\\ 
&+&\mathcal{Q}_1\mathcal{H}\mathcal{P}\frac{1}{\mathcal{P}(\omega-\mathcal{H}-i\epsilon)\mathcal{P}}\mathcal{P}\mathcal{H}\mathcal{Q}_1 \nonumber\\
&+& \mathcal{Q}_1\mathcal{H}\mathcal{Q}_2\frac{1}{\mathcal{Q}_2(\omega-\mathcal{H}-i\epsilon)\mathcal{Q}_2}\mathcal{Q}_2\mathcal{H}\mathcal{Q}_1 
\nonumber\\  
&+&\mathcal{Q}_1\mathcal{H}\mathcal{Q}_2\frac{1}{\mathcal{Q}_2(\omega-\mathcal{H}-i\epsilon)\mathcal{Q}_2}\mathcal{Q}_2\mathcal{H}\mathcal{P}\frac{1}{\mathcal{P}(\omega-\mathcal{H}-i\epsilon)\mathcal{P}}\mathcal{P}\mathcal{H}\mathcal{Q}_1 \nonumber\\
&+& \mathcal{Q}_1\mathcal{H}\mathcal{P}\frac{1}{\mathcal{P}(\omega-\mathcal{H}-i\epsilon)\mathcal{P}}\mathcal{P}\mathcal{H}\mathcal{Q}_2\frac{1}{\mathcal{Q}_2(\omega-\mathcal{H}-i\epsilon)\mathcal{Q}_2}\mathcal{Q}_2\mathcal{H}\mathcal{Q}_1\nonumber\\
\end{eqnarray}

\ack
Y. N. acknowledges partial support by the National Natural Science Foundation of China under Grant No. 11305161.

\section*{References}

\bibliography{bibliography}

\end{document}